\documentclass[aps,superscriptaddress]{revtex4-1}
\usepackage{graphicx, amsmath,subcaption}
\usepackage{color}

\newcommand{\be}{\begin{equation}}
\newcommand{\ee}{\end{equation}}
\newcommand{\bea}{\begin{eqnarray}}
\newcommand{\eea}{\end{eqnarray}}

\newcommand{\bel}[1]{\begin{equation}\label{#1}}
\newcommand{\beal}[1]{\begin{eqnarray}\label{#1}}

\begin{document}


\title{Resolution of cosmological singularity in Ho\v{r}ava-Lifshitz cosmology}


\author{Ewa Czuchry}
\email[]{ewa.czuchry@ncbj.gov.pl}
\affiliation{National Centre for Nuclear Research, Pasteura 7, Warsaw, Poland}


\date{\today}

\begin{abstract}

The standard $\Lambda$CDM model  despite its agreement with observational data  still has some issues unaddressed, lie the problem
of initial singularity. Solving that problem usually requires modifications of general
relativity. However, there appeared the
Ho\v{r}ava-Lifshitz (HL) theory of gravity, in which equations governing cosmological evolution include a new term scaling similarly as    dark radiation term in the   Friedmann equations, enabling  a bounce of the universe instead  of  initial singularity. This review 
describes past works on a stability of such a bounce in different formulations of  HL theory, initial detailed balance scenario and further projectable versions containing higher than quadratic term to the original action.

\end{abstract}


\maketitle

\section{Introduction}
Classical General Relativity (GR) apart its simple beauty and symmetry is also  strongly confirmed in several experimental tests. However, it does not explain many issues like dark matter, spacetime singularities including the initial one in cosmology, and the ones inside  the black holes.  In order to answer these  issues there has been many attempts  to modify GR both on the classical and quantum level.  Specifically, the quantisation of GR cosmology was supposed  to resolve the initial  singularity problem. 

Attempts to quantize gravity could be divided into two categories. One way was to assume the classical theory of gravity and quantise it in various manners, with the first attempts performed via  the  the covariant quantum gravity. In that classical approach one repeats the method successful in quantising electrodynamics, namely considering the path integral of the Hilbert-Einsteim action and then calculates the perturbation of  the metric around a background one. The obtained equations  unlike in electrodynamics are  non-renormalizable  in higher energies.
The canonical quantum gravity considers ADM $(3+1)$-decomposition of the spacetime and quantisation of the constraints obtained from Hamiltonian.  Other attempts included   sophisticated theories like string theory and loop quantum gravity. These theories manage to solve  some problems (such as a cosmological singularity \cite{PhysRevLett.92.071302}) but there are difficult to be  phenomenologically tested ~\cite{Quevedo:2016tbh,Girelli:2012ju}. There are also  attempts for resolving an initial singularity problem by combination of canonical and coherent state quantisation like the one in our  paper \cite{Bergeron:2015ppa}, however at this moment they are difficult to be validated by observations data.

 Although there is still no full theory of quantum gravity developed  it is  supposed to manifest beyond a characteristic energy scale for quantum gravity  $E_{Pl} = \sqrt{\hbar c^5/G}$ built in terms of the speed of light $c$, the gravitational constant $G$  and  the Planck's constant $\hbar$.
  Therefore, there is the second  research direction which aims to  construct a modified version of GR with  an improved UV behavior.  
General relativity after many tests performed seems to be consistent with  all current observations. This makes it  a very good IR limit  of potential quantum gravity model. There has been made some  proposals for UV-completions of general relativity in the past~\cite{Arkani-Hamed:1998jmv, Dvali:2007hz}. They  have one thing in common, namely the existence of  some cutoff energy scale beyond which quantum effects could be detected, specially for a cutoff energy in  the range of TeV. The widely discussed recent  proposal  is Ho\v{r}ava gravity, which is a proposal of a UV complete theory of gravity. It seems to be   renormalizable at high energies, which makes it a   candidate for a quantum gravity model~\cite{Horava:2009uw,Horava:2010zj}. 
 The action of this theory contains  additional   higher order spatial derivatives and therefore 
 the theory loses the full diffeomorphism invariance, keeping  the (1+3) foliation preserving diffeomorphism. Moreover, there is  an UV fixed point in this gravity model where there is  an anisotropic Lifshitz scaling between time and space. Therefore,  the resulting theory  is  called Ho\v{r}ava-Lifshitz (HL) gravity.

Significant work has been done on this theory where different aspects and  properties were examined~\cite{Audren:2014hza,Blas:2009yd,Blas:2011zd,Calcagni:2009ar,Colombo:2015yha,Czuchry:2009hz,Czuchry:2010vx,Dutta:2009jn,Dutta:2010jh,Frusciante:2015maa,Kiritsis:2009sh,Lu:2009em,Saridakis:2009bv,Sotiriou:2009bx}.   Many  studies were devoted to cosmological solutions~\cite{Mukohyama:2010xz,Kiritsis:2009sh,Calcagni:2009ar}, braneworlds and dark radiation~\cite{Calcagni:2009ar, Saridakis:2009bv}.
Ho\v{r}ava-Lifshitz cosmology obtained a novel feature  enabling the existence of bounce instead of initial singularity predicted by classical GR. There has been also other research focused on finding specific solutions, including black holes, and their properties, and many  works devoted to phenomenological aspects both astrophysical and concerning dark matter.

Derivation of Ho\v{r}ava-Lifshitz cosmology~\cite{Mukohyama:2010xz,Kiritsis:2009sh,Calcagni:2009ar} made via varying action written Friedmann-Robertson-Walker space-time metrics resulted in equations analogous to the standard Friedman ones.  These equations contain   a new term which scales similarly as
dark radiation \cite{Calcagni:2009ar,Kiritsis:2009sh,Saridakis:2009bv}, i.e.  $\sim 1/a^4$   (where $a$ is a scale factor),  and provides a negative
contribution to the energy density. This feature enables obtaining  non-singular cosmological evolution, resolving  the initial singularity problem \cite{Saridakis:2009bv,Brandenberger_2009,Czuchry:2009hz}. Such a
possibility  not only results in avoiding the initial singularity but may have other  consequences for potential histories of the Universe  like scenario of  contraction from the infinite size connected by a  bounce to the expansion to infinite size again, or eternal cycles of the similar scenario.

 Despite many promises made by this modified theory of gravity it seems that it contains
 instabilities and pathologies in different formulations  (see e.g.  \cite{Sotiriou:2009bx,Blas:2009qj, Sotiriou:2010wn,Wang:2017brl}). The original Ho\v{r}ava formulation suffers  from among many problems: the existence of  ghost instabilities and strong coupling at IR \cite{Blas:2009yd, Charmousis:2009tc}, the appearance  of a  term that violates parity \cite{Sotiriou:2009bx},
  very large value and negative sign of cosmological constant \cite{Vernieri:2011aa, Appignani:2009dy}, issues with power counting renormalisation of the propagation of the scalar mode  \cite{Vernieri:2015uma,Colombo:2015yha}.  Some of those problems  might be solved by  performing  an analytic continuation of the parameters of the theory \cite{Lu:2009em}. 
 
 In the original Ho\v{r}ava formulation it is assumed  via the so called  detailed balance condition  that a potential part of the action is derived from the so-called superpotential, which limits the big number of its terms and corresponding independent couplings. Another imposed condition is the demand of  projectability, used in a standard cosmology.  It requires  that  lapse function $N$ depends only
on time ${ N}={ N}(t)$.  It might seem that this condition is too strict but on the other hand  it seems that non-projectable version of  Ho\v{r}ava gravity  results in  serious strong coupling problem (\cite{Charmousis:2009tc})  and does not possess a valid  GR limit at IR. However, some authors \cite{Pospelov:2010mp, Vernieri:2011aa} claim the opposite, proposing adding additional terms to the superpotential (not to the action thus still keeping detailed balance or eventually softly breaking it)  and relaxing projectability.  Nonetheless, subsequent works demonstrated that it caused  problems with  the scalar mode power-counting renormalizability.

One of the simplest models with  the detailed balance condition  relaxed  is  the  Sotiriou-Visser-Weinfurtner (SVW) generalisation~\cite{Sotiriou:2009bx}. This version of HL gravity  assumes a gravitational action containing terms not only quadratic in curvature, but also cubic    ones,  what was suggested  already in \cite{Calcagni:2009ar,Kiritsis:2009sh}. This model still maintains the projectability condition. Generalised Friedmann equations obtained from varying such an action  contain not only a 
 dark radiation term  $\sim 1/a^4$  but also terms scaling as $\sim 1/a^6$ term. These new terms, although  negligible at large values of $a$, become dominating at small ones and might modify or cancel bounce solutions. Specifically, as it has the opposite sign than the $1/a^4$ term, it may compensate the dark radiation term at small scales and result in singular solutions. Similar scenario arrives  in the HL gravity with the softly broken  detailed balance condition and negative spacial curvature \cite{Son:2010ci}.

Nonetheless, the issue of the initial singularity still
remains one of the key questions of early Universe cosmology. The possibility that it might be  avoided in a modified gravity and replaced by a bounce is a very promising feature.
In this review we are going to present result of the research \cite{Czuchry:2009hz,Czuchry:2010vx} performed via phase portrait techniques, on occurrence and stability of the bounce in two simplest formulation of HL cosmology: original one with imposed detailed balance condition and SVW formulation relaxing this  condition.  As additional terms in analogs of Friedman equations are proportional the curvature parameter $K=\{-1,0,1\}$ only non-flat cosmologies with $K=\pm1$ allow the existence of a bounce and existence of  non-singular  solutions.

In \cite{Czuchry:2009hz} matter sector was described  in terms of a scalar field with a  potential given by a quadratic power of that field. More general approach and easier for fitting with observational data is  the  hydrodynamical approach used in \cite{Czuchry:2010vx}  where matter sector is described in terms of  density $\rho$ and  pressure $p$.  In the latter work it was assumed that  $w$ providing the relation between density and pressure in the equation of state,  is constant, which  is at some level an idealisation and simplification. At the moment  we do not have the  history of the  HL universe constructed in a similar way as in the standard   $\Lambda$CDM model, where we have   phases and epochs containing  different matter or radiation sectors. Therefore, as we still have   limited  understanding on the physical aspects of the theory and its parameters, current  research  rather describes    different analytical possibilities, not some exact physical solutions.

This paper is organised as follows: We first give a brief overview  of HL cosmology in both scenarios under consideration in Section II.  In Section III the possibility of
bounce in both formulations is discussed. Section IV contains derivation and description of the phase portraits of the HL cosmology with imposed condition of detailed balance, while in the section V this condition is released.  Section VI  contains summary of results on possibility of a bounce in HL cosmology. In Section VII we discuss limitation of the underlying  theory.

\section{Ho\v{r}ava-Lifshitz cosmology}
The main obstacle in quantising gravity is  that general relativity in its classical formulation is non-renormalisable.  This might be visualised by  expanding some quantity $\mathcal{F}$ with respect to   the gravitational constant~\cite{Wang:2017brl} as follows:
\begin{equation}
\mathcal{F} = \sum_{n=0}^\infty  a_n\left(G_N E^2\right)^n.
\end{equation}
Here $E$ is the energy of the system, $a_n$ denotes  a numerical coefficient and $G_N$ is the gravitational coupling constant.
Therefore,  $E^2 \geq G^{-1}$ and the expansion  above diverges. Consequently, as demonstrated, general relativity is not  perturbatively renormalisable in the  high energy regimes.

There has been many researches pointing out that  the ultraviolet behaviour of general relativity might be improved  by including higher-order derivatives in the standard gravitational metric. The latter is the Einstein-Hilbert action:
\be
S = \int d^4x \sqrt{g}R,
\ee
where  d$^4x$ denotes  volume element of the space-time, $g$ is  its metric matrix' determinant  and $R$ is a scalar curvature. Including higher order terms of the derivatives of the metric  provides  a following action:
\be
S = \int d^4x \sqrt{g}(R+f(R_{\mu\nu}R^{\mu\nu})).
\ee
The additional terms, containing different derivatives of $R$, $R_{\mu\nu}$ etc.  change the graviton propagator from $1/k^2$ into $1/(k^2-G_Nk^4)$~\cite{Horava:2009uw,Horava:2010zj}. The propagator part   proportional to $k^{-4}$ cancels the ultraviolet divergence. However, the resulting theory has time derivatives of $\mathcal{O}>2$ and therefore   non-unitary. Moreover,  it possesses a  spin-2 ghost with a non-zero mass~\cite{Wang:2017brl} and derived form that action field equations are of the fourth order. 

The novel idea of  Ho\v{r}ava~\cite{Horava:2009uw} was to construct a higher-order theory of gravity breaking the Lorentz invariance in the ultraviolet. In his theory only the \emph{spatial} derivatives are of $\mathcal{O}>2$ which evaded the ghost. However,  it is demanded that any theory of gravity theory  should be consistent with all current experiments which have not  detected any significant violation of Lorentz invariance. Thus it is necessary to restore the   Lorentz invariance  in the infrared limit. In order to overcome this problem Ho\v{r}ava  proposed  an anisotropic scaling of space and time in high UV energies, which is  known as Lifshitz scaling. In a 4-dimensional spacetime this scaling takes the form:
\begin{equation}
t \to b^{-z}t, \, x^i \to b^{-1}x^i,
\end{equation}
here $\, i=1,2,3$ and $z$ is a critical exponent. Lorentz invariance is restored when $z=1$, but the power-counting renormalisability demands $z\geq 3$~\cite{Wang:2017brl}, usually $z=3$ is assumed. Therefore, the resulting theory is called Ho\v{r}ava-Lifshitz (HL) gravity.   Lorentz symmetry  is here  broken down to transformations $t \to \xi_0(t), \, x^i \to \xi^i(t,x^k)$, preserving the spatial diffeomorphisms unlike full space time diffeomorphisms invariance of GR. Thus such a theory acquires a symmetry   preserving a space-time  foliation~\cite{Horava:2009uw,Wang:2017brl},  where on each constant time hypersurface  there are allowed arbitrary changes of the spatial coordinates.

Preservation of  a space-time  foliation  and anisotropic scaling between time and space and time  introduces the ADM (1+3)decomposition of the spacetime. The  standard ADM metrics  in a preferred foliation and  with $(-+++)$ signature is following:
\begin{equation}
\text{d}s^2 = -N^2\text{d}t^2 + g_{ij}(\text{d}x^i+N^i\text{d}t)(\text{d}x^j+N^j\text{d}t).
\end{equation}
The dynamics  is now described in terms of  the lapse function $N$, the shift vector $N^i$, and the spatial metric $g_{ij}$ ($i$, $j=1,2,3$). The most general action for such theory can be written as:
\begin{equation}\label{eq:gen_action}
S = \int \text{d}^3x\text{d}t\ N\sqrt{g}\left[K^{ij}K_{ij}-\lambda K^2-\mathcal{V}(g_{ij})\right].
\end{equation}
Here as usually $g$ denotes the determinant of the spatial metric $g_{ij}$, $\lambda$ is a dimensionless running coupling constant,  $\mathcal{V}$ is a potential term and  $K$ is a trace of the extrinsic curvature of the spatial 3-dimensional hypersurface $K_{ij}$:
\begin{equation}
K_{ij} = \frac{1}{2N}\left(\dot{g}_{ij}-\nabla_iN_j-\nabla_jN_i\right).
\end{equation}
An overdot denotes a derivative with respect to the time coordinate $t$. The trace of $K_{ij}$ is  $K$. The potential $\mathcal{V}$ is  invariant only under three-dimensional diffeomorphisms~\cite{Blas:2009qj} and depends only on the spatial metric and its (spatial) derivatives. Thus it  contains only operators constructed from the spatial metric $g_{ij}$ and of dimension 4 and 6.

\subsection{Detailed Balance}
As the action (\ref{eq:gen_action}) is  very complicated  Ho\v{r}ava~\cite{Horava:2009uw,Sotiriou:2010wn,Vernieri:2011aa} proposed to impose additional condition, the so-called detailed balance.  It  assumes that the $\mathcal{V}$  could be derived from a superpotential $W$~\cite{Horava:2009uw,Sotiriou:2010wn,Vernieri:2011aa}:
\begin{equation}
\mathcal{V} = E^{ij}\mathcal{G}_{ijkl}E^{kl}, \quad E^{ij} = \frac{1}{\sqrt{g}}\frac{\delta W}{\delta g_{ij}},
\end{equation}
and
\begin{equation}
\mathcal{G}^{ijkl} = \frac{1}{2}\left(g^{ik}g^{jl}+g^{il}g^{jk}\right) - \lambda g^{ij}g^{kl}.
\end{equation}
By carrying out    an analytic  continuation (e.g. \cite{Lu:2009em}) of two constant parameters $\omega$ and $\mu$ we obtain he action for Ho\v{r}ava-Lifshitz gravity in the detailed balance condition~\cite{Sotiriou:2010wn}   and reads as
\begin{equation}\label{eq:action}
\begin{aligned}
S_{db} = \int \text{d}t\, \text{d}^3x\sqrt{g}N&\Bigg[\frac{2}{\kappa^2}\left(K_{ij}K^{ij}-\lambda K^2\right)+\frac{\kappa^2}{2\omega^4}C_{ij}C^{ij}-\frac{\kappa^2\mu}{2\omega^2}\frac{\epsilon^{ijk}}{\sqrt{g}}R_{il}\nabla_jR^l_k \\&+\frac{\kappa^2\mu^2}{8}R_{ij}R^{ij}+\frac{\kappa^2\mu^2}{8(1-3\lambda)}\left(\frac{1-4\lambda}{4}R^2+\Lambda R -3\Lambda^2\right)\Bigg],
\end{aligned}
\end{equation}
where $C^{ij}$ is the Cotton tensor:
\begin{equation}
C^{ij}=\epsilon^{ikl} \nabla_k \left (R^j_{\ l}-\frac{1}{4}R\delta^j_l\right),
\end{equation}
 $\epsilon^{ikl}$ denotes the totally antisymmetric tensor. The parameters $\kappa, \omega$, and $\mu$ arriving in the theory have mass dimension respectively $-1$, $0$, and $1$.  The  analytic  continuation mentioned above  reads as $\mu \mapsto i\mu$ and $\omega^2\mapsto-i\omega^2$  and it enables obtaining the positive values of  the cosmological constant $\Lambda$ as  predicted by  current observational results in the low energy regime.

It is expected that action (\ref{eq:action}) reduces  to the Einstein-Hilbert one  in  the IR limit of the theory.
This is possible if  the speed of light $c$ and gravitational constant $G$ correspond to  HL parameters as follows:
\begin{equation}\label{eq:cosmoconstants}
 G= \frac{\kappa^2}{32\pi c},\quad c = \frac{\kappa^4\mu^2\Lambda}{8(3\lambda-1)^2} .
\end{equation}

The coupling constant $\lambda$ present in the action \eqref{eq:action} is dimensionless. It runs  with energy and  flows to the three infrared (IR) fixed points (\cite{Horava:2009uw}): $\lambda=1/3$, $\lambda=1$ or $\lambda=\infty$. However, some of those values seem unphysical, in the region $1>\lambda>1/3$ there appear ghost  instabilities in the IR limit of the theory \cite{Bogdanos:2009uj}. The attempt to solve this  problem \cite{Lu:2009em} resulted in instabilities re-emerging at the other energy region, in UV.  Thus 
the most  physically interesting case is the regime $\lambda\ge1$ that 
allows for a possible flow towards GR, where  $\lambda=1$. Region $\lambda\le 1/3$ on the other hand is disconnected from $\lambda=1$, therefore cannot be included in realistic physical considerations.

In order to obtain a  cosmological model  it is necessary to populate the universe with  matter (and radiation). The simplest method would be to  model the matter sector by assuming it is described by a scalar field $\varphi$ with a quadratic potential $V(\varphi) =  \frac{1}{2} m^2 \varphi^2 $. However,  a more realistic approach is to apply 
  a hydrodynamic approximation where matter is described by two quantities  $p$ and $\rho$, which are respectively pressure and energy density and fullil the continuity equation $\dot{\rho}+3H(\rho+p)=0$.

To derive equations of HL cosmology  one uses  the projectability condition $N=N(t)$ \cite{Horava:2009uw} and the spatial part of the metrics being the standard FLRW line element: $g_{ij} = a^2(t)\gamma_{ij},\, N_i=0$, where $\gamma_{ij}$ denotes a
maximally symmetric  metric with constant curvature:
\begin{equation}\label{eq:friedmann0}
\gamma_{ij}\text{d}x^i\text{d}x^j = \frac{\text{d}r^2}{1-Kr^2}+r^2(\text{d}\theta^2+\sin^2\theta\text{d}\varphi^2),
\end{equation}
values   $K=\{-1,0,1\}$ correspond respectively to closed, flat, and open Universe.
This background metric implies that
\begin{equation}
C_{ij}=0\,,\qquad R_{ij}=\frac{2K}{a^2}g_{ij}\,, \qquad K_{ij}=\frac{H}{N}g_{ij}\,, 
\end{equation}
where $H\equiv \dot a/a$ denotes the Hubble parameter.

On this background the  gravitational action (\ref{eq:action}) take the following form :
\be
S_{\rm FRW}=\int dt\, d^3x \, N a^3\,\left\{\frac{3(1-3\lambda)}{2\kappa^2}\frac{H^2}{N^2}+\frac{3\kappa^2\mu^2\Lambda}{4(1-3\lambda)}\left(\frac{K}{a^2}-\frac{\Lambda}{3}\right)
-\frac{\kappa^2\mu^2}{8(1-3\lambda)}\frac{K^2}{a^4}\right\}.\label{hla3}
\ee
In order to obtain  equations of motion on a cosmological background  one needs to vary the action \eqref{hla3}   with
respect to $N$ and  $a$. Only after that and  the lapse can be set  to one: $N = 1$  and  terms with  density $\rho$ and pressure $p$ are added. This procedure provides the analogs to  the Friedmann equations for the projectable Ho\v{r}ava-Lifshitz cosmology with imposed the detailed-balance condition:

\bea
H^2 &=&\frac{\kappa^2 \rho}{6(3\lambda-1)} \pm \frac{\kappa^4\mu^2}{8(3\lambda-1)^2} \left( \frac{ K\Lambda}{a^2} - \frac{\Lambda^2}2-\frac{ { K}^2 }{2a^4}\right)
, \label{eq:friedmann1} \\
{\dot H}  &=& -\frac{\kappa^2(\rho+p)}{4(3\lambda-1)} \mp \frac{\kappa^4\mu^2}{8(3\lambda-1)^2}\left( \frac{K\Lambda}{a^2} +  \frac{ { K}^2 }{4a^4} \right), \label{eq:friedmann2}
\eea
together with the  continuity equation:
\be
\dot{\rho}+3H(\rho+p)=0.\label{ce}
\ee
In the equations above there are two signs before the terms with  $\Lambda$, namely  the upper one corresponds the $\Lambda<0$ case, the lower one describes the analytic continuation $\mu\mapsto i\mu$ providing a positive $\Lambda$.

Some terms in  the above equations which scale as $a^{-4}$ are similar to  the dark energy expressions therefore  parameters: energy density $\rho_{de}$ and pressure density $p_{de}$  are interpreted as dark energy parameters:
 \begin{align}
\rho_{de}|_{db}&:=\frac{3\kappa^2\mu^2 K^2}{8(3\lambda-1)a^4}+\frac{3\kappa^2\mu^2\Lambda^2}{8(3\lambda-1)}\label{eq:de},\\
p_{de}|_{db}&:=\frac{\kappa^2\mu^2 K^2}{8(3\lambda-1)a^4}-\frac{3\kappa^2\mu^2\Lambda^2}{8(3\lambda-1)}.
\end{align}

We require that eqs~(\ref{eq:friedmann1}) and (\ref{eq:friedmann2}) coincide with the standard Friedmann equations. Thus, we can identify the following:\begin{equation}\label{eq5}
c=\frac{\kappa^2\mu}{4}\sqrt{\frac{\Lambda}{1-3\lambda}}, \ \
G=\frac{\kappa^2 c}{32\pi}, \ \  \Lambda_{E}=-\frac{3\kappa^4\mu^2}{3\lambda-1}\frac{\Lambda^2}{32}=\frac{3c^2}2 \Lambda,
\end{equation}
respectively, as well as $\mu^2\Lambda = 1/32\pi^2G^2$ and $\lambda = 1$ (which is an IR fixed point).  We demand  a real value of the speed of light $c$, therefore the  cosmological constant $\Lambda$ has to be negative for $\lambda>1/3$ and positive for $\lambda<1/3$. In order to obtain a positive cosmological constant $\Lambda$, as suggested by observations, it is necessary to perform in (\ref{eq:action}) an analytic complex continuation of constant parameters  $\mu$ and $\omega$  as follows $\mu \mapsto i\mu$ and $\omega^2\mapsto-i\omega^2$. On the level of equations for Ho\v{r}ava-Lifshitz cosmology varying $\lambda$-parameter  in the range $[1,\infty)$ results in   the running of the speed of  light, but does not change the structure of the equations  (\ref{eq:friedmann1}) and (\ref{eq:friedmann2}). 

When we substitute the equation of state $p=w\rho$ and  the above expressions  linking physical constants and HL parameters to  (\ref{eq:friedmann1}) and (\ref{eq:friedmann2}) we obtain the following equations:
\bea
H^2 &=& \frac2{3\lambda-1}\left[\frac{\rho}3\pm\left(\frac{\Lambda_{E}}{3}-\frac{K}{a^2}+\frac{3}{4\Lambda_{E}}\frac{K^2}{a^4} \right)\right]\label{hc11}\\
\dot{H}&=&\frac{2}{3\lambda-1}\left[-\frac{(1+w)}2\rho\pm\left(\frac{K}{a^2}-\frac3{2\Lambda_{E}}\frac{K^2}{a^4}\right)\right].\label{hc22}
\eea

\subsection{Beyond detailed balance}
The gravitational action \eqref{eq:action} contains terms up to quadratic in the curvature. However, a more  general renormalizable theory could  also  contain cubic terms and there is not {\it a priori} reason to keep only quadratic terms (\cite{Calcagni:2009ar,Kiritsis:2009sh,Wang:2017brl}).  Thus Sotiriou, Visser and Weinfurtner  (\cite{Sotiriou:2010wn})  built a  projectable theory as the  original Ho\v{r}ava theory, but without imposing the detailed balance condition in the action.

This formulation led to Friedmann equations with an additional term $\sim1/a^6$, moreover with additional  and uncoupled coefficients:
 \bea
H^2 &=& \frac2{(3\lambda-1)}\left(\frac{\rho}3 + \sigma_1 + \sigma_2  \frac{K}{a^2} + \sigma_3 \frac{K^2}{a^4} + \sigma_4 \frac{K}{a^6}\right),\\
\dot{H}&=& \frac2{(3\lambda-1)}\left(-\frac p2 - \frac\rho2 - \sigma_2\frac{K}{ a^2} -2\sigma_3\frac{K^2}{ a^4} -3\sigma_4\frac{K}{ a^6} \right).\
\eea
 In order to coincide with the Friedmann equations in the IR limit $\lambda=1$ and for large $a$, when  terms proportional to  $1/a^4$ and  to $1/a^6$ become negligibly small, one has to set $\sigma_1=\Lambda_{E}/3$ and $\sigma_2=-1$.  However, values of constants $\sigma_3$, $\sigma_4$ are at this stage arbitrary. This way we obtain the following equations:
 \bea
H^2 &=& \frac2{(3\lambda-1)}\left(\frac{\rho}3 + \frac{\Lambda_{E}}{3} -  \frac{K}{a^2} + \sigma_3 \frac{K^2}{a^4} + \sigma_4 \frac{K}{a^6}\right)\label{h2ndb},\\
\dot{H}&=& \frac2{(3\lambda-1)}\left(-\frac{\rho(1+w)}2 + \frac{K}{ a^2} -2\sigma_3\frac{K^2}{ a^4} -3\sigma_4\frac{K}{ a^6} \right)\label{hdndb},\
\eea
We can observe new terms in the above analogs of Friedmann equations, proportional to $1/a^6$. They  mimic stiff matter,  such that $\rho=p$ ($w=1$) which scales similarly  $\rho_{\textrm{stiff}}\sim1/a^6$). These terms are negligibly small at large scales, but may play a significant role at small values of a scale parameter, thus changing the dynamics of the Universe around initial singularity or a bounce.

\section{Existence of bounce}
Ho\v{r}ava-Lifshitz cosmological equations contain additional   terms  proportional $a^{-4}$ (DB) and to $a^{-6}$ (BDB) that introduce the possibility of a cosmological bounce, namely a scenario in which contraction of the universe stops and reverse to expansion (or in the opposite direction).  In a DB scenario, from the form of  eq. (\ref{eq:friedmann1}) it follows   that it is possible that $H=0$.
When this condition is fulfilled at some monet of time  the realisation of the bounce is possible (but not necessary, for that we also need $\dot{H}\neq 0$. In the case $\lambda=1$  \cite{Calcagni:2009ar}, the bounce may happen in non-empty Universe equipped with matter, at the critical
time $t_*$, $a=a_*$, when the critical energy density reaches the following value:
\be
\rho=\rho_*= \frac{12}{\kappa^2}\left(\frac{K}{a_*^2}+ \frac{\Lambda_E}{3}
+\frac{\kappa^4\mu^2}{64}\frac{ { K}^2}{ a_*^4}
\right),
\ee
This value  is determined by the values of couplings $\kappa$ and $\mu$.

Additionally, a  continuity
equation implies that at the bounce $\dot H> 0$. Therefore, when the condition $H=0$ is fulfilled we also have the sufficient condition for existence of a bounce $\dot{H}\neq 0$. As  $\dot H> 0$ is is only possible  a transition from a contracting to
an expanding phase, but not the reverse. Moreover, there is another  condition for a realisation 
bounce \cite{Brandenberger_2009}   that  requires that  $(\frac{\rho}{12} - p)  >  0$ and  the energy density of
regular matter scales  less fast than dark matter  terms.

Near the bounce so for small $a$ the dominating terms in the Friedmann equations  (\ref{hc11}) and (\ref{hc22})
 are  the terms scaling as $a^{-4}$,  while others  terms  become insignificant. 
 Particularly,   $H^2$, $\dot{H}$ and $\rho$
  scale as $a^{-3(1+w)}$, where  $w$ is a constant parameter in the   equation of state  $p = w
  \rho$.  Subsequently, if $w>-\frac13$ the density term dominates over  the
  curvature term $\sim1/a^2$.  
  
 In the BDB scenario bounce might happed at the critical density:
 \be
\rho_*= -\Lambda_E +3\frac{K}{ a_*^2} -3  \frac{\sigma_3K^2}{ a_*^4} -   \frac{3\sigma_4K}{a_*^6}.
\ee
For flat universe and positive cosmological constant bounce is not positive as resulting critical density becomes negative.    
\section{Bounce stability in the detailed balance formulation}

We are mainly interested in the possibility of appearing  of a bounce which could be given by dynamics of variables $a$ and $H$.  
From eq. \eqref{hc11} we might determine $\rho$ and then insert its formula into \eqref{hc22}. This way we obtain two systems, one containing the formula fo density and its derivative via the continuity equation \eqref{hc11}, but still dependent on $a$ and $H$. The second system is independent and consist of two equations  describing the evolution of  $a$ and $H$.

Specifically,   eq. \eqref{hc11} provided a following  expression for $\rho$:
\be
\rho=\frac{3(3\lambda -1)}2H^2\mp\left(\Lambda_E-3\frac{K}{a^2}+\frac9{4\Lambda_E}\frac{K^2}{a^4}\right).
\ee
This expression substituted in  \eqref{hc22} results in
\be
\dot{H}=\frac{\pm1}{3\lambda-1}\left[\left(1+w\right)\Lambda_E-\left(3w+1\right)\frac{K}{a^2}+\frac{3\left(3w-1\right)}{4\Lambda_E}\frac{K^2}{a^4}\right]-\frac32\left(1+w\right)H^2\label{pe1}.
\ee
Adding the   the definition of the Hubble parameter:  
\be
\dot{a}=aH\label{pe2},
\ee  
we have a two dimensional dynamical system.

The set of equations (\ref{pe1}-\ref{pe2}) is difficult to solve analytically. However, we are interested non in detailed solutions but in the qualitative analysis.
In purpose of that we use the method of the phase portraits, where we search for critical points and analyse their character.
These  points
are locations where the
derivatives of all the dynamic variables, In our case when  the r.h.s.  of
(\ref{pe1}-\ref{pe2}), vanish.  What we obtain are   the only points where phase
trajectories could  start, end or intersect.  Moreover, they can also appear in
infinity. In this case  a suitable coordinate transformation, the so called Poincar\'{e} projection, is used that projects the
complete phase space onto a compact region. The  nature  of these points, both finite and infinite,  is given by the properties of
the Jacobian matrix of the linearized equations at those points. All that information  provides a
qualitative analysis  of the  dynamical system.

The method of  finding   critical points  consists of setting  all right-hand-sides of
dynamical equations to zero, thus finding points where derivative of dynamical variables vanish. In case of two equations  \eqref{pe1}-\eqref{pe2}  the corresponding solutions are two following $P_1$ and $P_2$ in phase-space $(a,H)$:
\bea
P_1: a^2&=&\frac{3K}{2\Lambda_E},\ H=0,\\
P_2: a^2&=&\frac{(3w-1)K}{(1+w)2\Lambda_E},\ H=0.
\eea
These two points exist when the values of $a$ obtained via square root of the expression on the right hand side of the above equations are real and nonnegative. Thus the point $P_1$  exists if $K/\Lambda_E>0$, if we assume a positive cosmological constant therefore only for $K>0$. Point $P_2$  exists when $w>1/3$ and $K/\Lambda_E>0$ or $w<1/3$ and $K/\Lambda_E<0$. Thus  we have  two critical points existing   if the parameter of state  $w>1/3$. Moreover, they  are both finite, unless $w=-1$, when  $P_2$ blows to  infinity.  As mentioned above, due to $\dot H> 0$ both points represent a bouncing solution.

In order to complete the analysis, the  stability properties of the critical points needed.
 They are determined by the eigenvalues of the Jacobian $A$ of the system
\eqref{pe1}-\eqref{pe2}.  Eigenvalues of $A$  with non-zero real parts different from zero point to hyperbolic points. They include sources (unstable) with positive real parts, saddle for
real parts of opposite  sign and sinks (stable) corresponding to  negative real parts. Critical points at which all the eigenvalues have
real parts different from zero are called hyperbolic. Among them  one can distinguish  sources (unstable) with positive real parts, saddles with
real parts of different sign and sinks (stable) for negative real parts. If at least one eigenvalue has a real part equal to
zero it is then called a non-hyperbolic critical point. For such points  it is not
possible to obtain conclusive information about the stability
from the Jacobian matrix  and other tools like
e.g. numerical simulation \cite{arrowsmith1990introduction} should be then used.

In the case of \eqref{pe1}-\eqref{pe2} the eigenvalues of $A$ at $P_1$ are both imaginary  and it is a center for all the values of the parameters.
The character of $P_2$  is more complicated and depend on the values of  $\Lambda_E$, $K$ and $w$. Thus $P_2$ is a center when $K/\Lambda_E<0$ and $-1\le w<1/3$, with a special subcase for $w=-1$ that being so it becomes a linear center, so a center with only one eigenvector. Otherwise it becomes a saddle, thus without a bouncing possibility.

To have a full picture of the dynamics of the Universe also the information about   critical points that occurring  at infinity is necessary. For this purpose  the so called Poincar\'{e} projection \cite{felder2002warped,felder2002cosmology} is used. It   projects the
whole infinite  phase space  $(a,H)$ onto a compact region. Specifically, we introduce  the new coordinates $(\tilde{a},\tilde{H})$ which  written in polar coordinates ${r,\phi}$: $\tilde{a}=r\cos\phi$ and $\tilde{H}=r\sin\phi$. Moreover:
\bea
a &=& {{r} \over 1-r} \cos\phi , \label{poin1} \\
H &=& {{r } \over 1-r}\sin\phi,  \label{poin2} 
\eea
It is also necessary to  rescale the
time parameter  $t$, which take infinite values,   introducing  the new time parameter $T$ in a similar way, i.e. $d{T} = dt/(1-r)$. In such
coordinates the  phase space  is now compactified to  a sphere of radius one and its interior.
Here infinity corresponds to $r=1$. We have to keep in mind that  a scale factor $a$ may take only nonnegative values, so actually a semi-sphere. 

This procedure provides the dynamical equations in terms of $r$, $\phi$ and their derivatives with respect to new time $T$. 
At the surface of the sphere, so limit $r = 1$ there are 3 solutions $P_3=(1,0)$, $P_4= (1,\pi/2)$, $P_6=(1,-\pi/2)$ written
in polar coordinates $(r,\phi)$. These critical points are now hyperbolic unless $w=-1$, resulting   in  $P_4$ and $P_6$ being respectively a repelling  and an attracting node.  For $w=-1$ points $P_4$ and $P_6$ are non-hyperbolic and  numerical simulations provide  that they are  saddles  and also  ends of a separatrice. 

The numerical phase portraits are presented at fig. 1, which  contains the deformed phase space, scaled to fit on the compactified sphere. We observe that bounce scenarios are only  possible when  one of the critical points  $P_{1}$ and $P_{2}$   exist and is a  center. Then we have  closed orbits around them and the Universe might go through eternal cycles of  expansion and collapse, connected by a bounce of a finite size, expansion etc.  However,  the point $P_{1}$ describes less physical bouncing solution with the density  $\rho=0$. More interesting case is when  $P_{2}$ is a center,  as a density $\rho$ is non-zero at that point. The special case of  $w=-1$ provides a  third bounce scenario is around the  linear center $P_{2}$ located now at  $\infty$. In this case  the universe begins in a  static infinite state as $H=0$ $a=\infty$, then contracts to  a finite size and rebounces to  a static infinite universe.

\begin{figure}
\includegraphics[height=75mm]{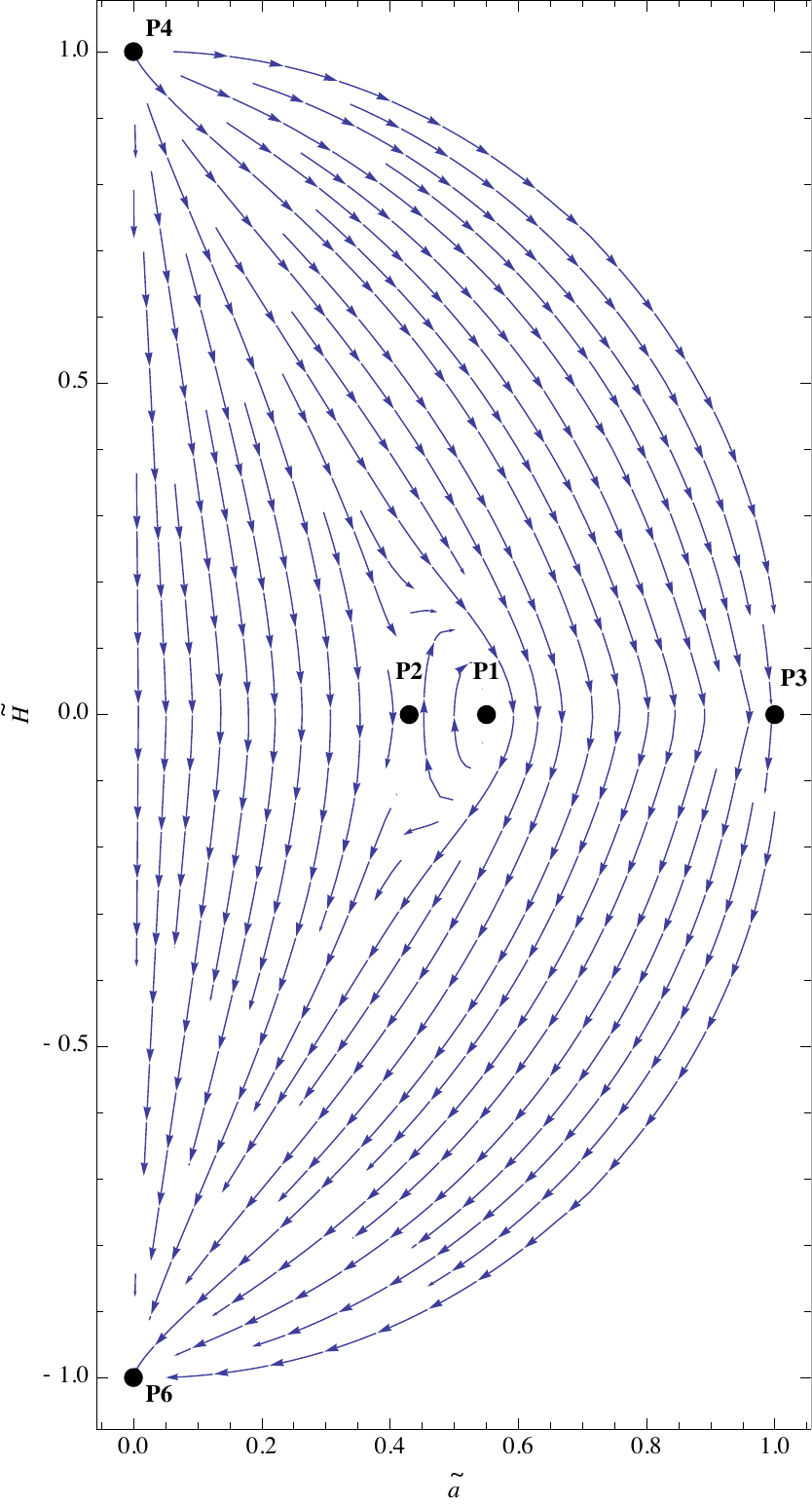}\includegraphics[height=75mm]{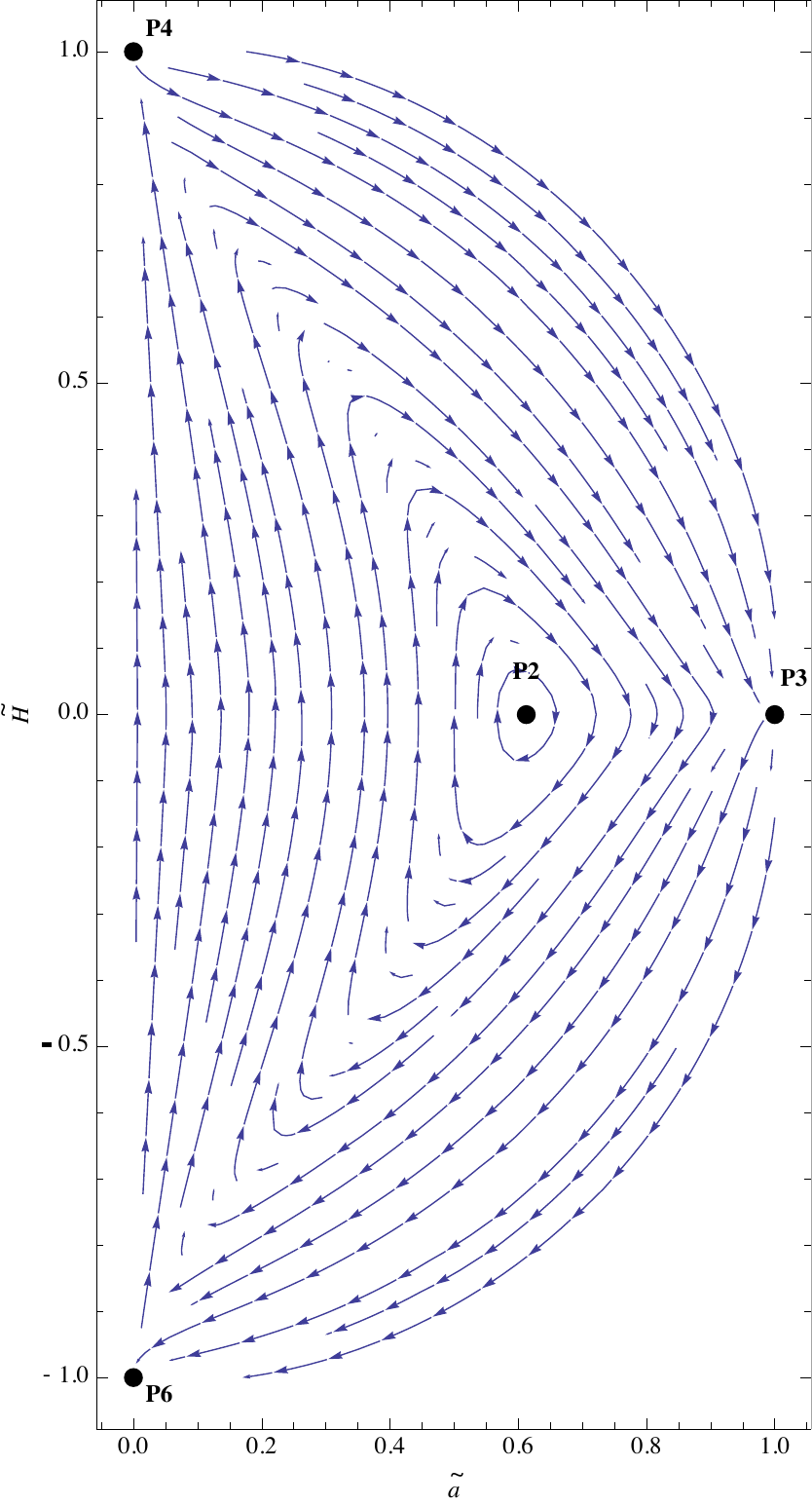}
\caption{Projected phase space of HL universe \cite{Czuchry:2010vx} in DB condition. On the left a case with $K\Lambda_E>0$ and $w>1/3$, on the right  $K/\Lambda_E<0$ and $-1<w<1/3$.}
\end{figure}

\section{Bounce stability in the beyond detailed balance formulation}
In the Sotiriou, Visser and Weinfurtner formulation the generalised Friedmann equations \eqref{h2ndb}-\eqref{hdndb} contain additional terms $\sim1/a^6$ and uncoupled coefficients.

Solving eq. \eqref{h2ndb} for $\rho$ provides:
\be
\rho=3 \frac{(3\lambda-1)}2 H^2 -\Lambda_E - 3\frac{K}{ a^2} -3  \frac{\sigma_3K^2}{ a^4} -   \frac{3\sigma_4K}{a^6}.
\ee
Substituting this expression on $\rho$ into (\eqref{hdndb}) and using the equation of state $p=w\rho$ results in
\be
\dot{H}=  \frac2{3\lambda-1}\left(   \frac{\Lambda_E(1 + w)}{2} -  \frac{ K(1+3w)}{2 a^2}  +\frac{\sigma_3 (-1 + 3 w)K^2}{2 a^4}
+\frac{ 3\sigma_4 (1 +  w)K}{2 a^6}\right)-\frac{3(1 + w)}{2} H^2 \label{doth}.
\ee
As in the DB case, supplementing  the above equation with the definition of the Hubble parameter provides the two dimensional dynamical system.

Again we search for critical  points where $\dot{a}$ and $\dot{H}$  these points fulfil $H=0$ and obtain the following condition:
\be
 \Lambda_E(1 + w)a^6 -  { K(1+3w)}{ a^4} + {\sigma_3 (-1 + 3 w)K^2}{ a^2}
+{ 3\sigma_4 (-1 +  w)K}=0\label{bic}
\ee
It  is a bicubic equation, which in general possesses quite complicated solutions but  might be simplified in two special cases. Namely when  $w=-1$ describing  the equation of state of the cosmological constant, and in case of radiation described by $w=1/3$.
Besides these two cases critical points of the system  (\ref{pe2}) and (\ref{doth})  have following coordinates $(a_x,0)$. Here $a_x^2$ is a root of the cubic equation:
 \be
 \Lambda_E(1 + w)x^3 -  { K(1+3w)}{ x^2} + {\sigma_3 (-1 + 3 w)K^2}{ x}
+{ 3\sigma_4 (-1 +  w)K}=0\label{bic1}.
\ee
  Such an equation might have zero, one, two or three real solutions depending on the sign  of its discriminant. Moreover, if they exist they are  either always stable or always unstable
depending on the sign of $K/(3\lambda-1)$.  Their character  depends on the values of $a_x$, $\Lambda_E$, $\sigma_3$ and $\sigma_4$. The most significant feature of oscillating (and bouncing) solutions in the SVW formulation is the existence of two  centres, with a saddle between them (three finite critical points) for some values of parameters.  In a more realistic situation, that includes dynamical change of state parameter, it would be possible to go from one oscillating  bouncing solution to another.

In order to study stability properties of infinite critical points one again has to perform 
 the Poincar\`e transformation. It leads to the similar results as in  detailed balance scenario.
Points at  infinity are transformed to the sphere  $r=1$. Two points at  $\phi=\pi/2$ and at $ -\pi/2$ are respectively repelling  and attracting  node, respectively. The point at  $\phi=0$ is non-hyperbolic.

Figure 2. shows the example of the phase space of system with three  finite critical points. Here  points $S_{1}$ and $S_{3}$ are centres and a point $S_{2}$ is a saddle.
\begin{figure}[!h]
\includegraphics[height=75mm]{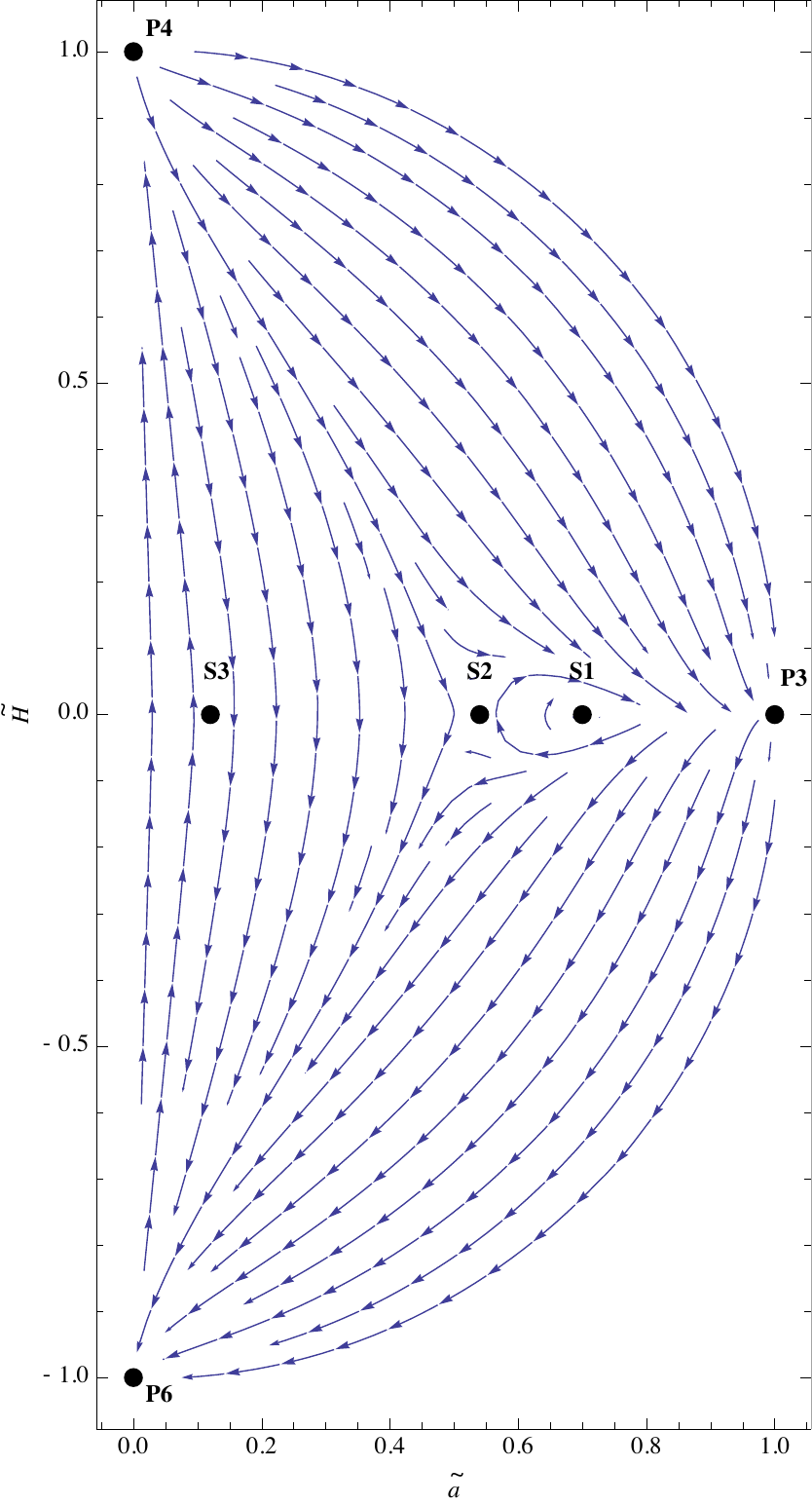}
\caption{Projected phase space of the HL universe in beyond detailed balance formulation with 3 critical points existing\cite{Czuchry:2010vx}.}
\end{figure}

\section{Discussion}

This paper reviews the research performed  the cosmological bounce in different formulations of projectable versions of 
Ho\v{r}ava-Lifshitz gravity, with and without detailed balance condition. The analogs of the Friedmann equations in
both this models contain a  term scaling as $1/a^4$ and similar to dark radiation.  That additional term enables that fthe Hubble parameter might be $H=0$ at some moment of time. This  is a necessary  condition for the realisation of the bounce while   an additional condition $\dot{H}\neq0$ makes it sufficient.   In the Sotiriou, Visser and Weinfurtner  model there is an additional term $1/a^6$ in the analogs of Friedmann equations. This term is of arbitrary sign, so it can enhance the possibility of a bounce or cancel it. 
 
 The biggest difference between the detailed balance theory and its breaking arrives for the small values of a scale parameter  $a$  as  the SVW gravity term $1/a^6$  plays role only  for the small values of $a$ and becomes insignificant for the bigger ones. This difference is visible in phase portraits of both theories and number of potential bouncing solutions. In the original Ho\v{r}ava formulation there exists one bouncing solutions for all values of parameters but it corresponds to density 
 $\rho=0$. For non zero  $\rho=0$ there might be a bouncing solution if  $K/\Lambda_E<0$ and $-1\le w<1/3$,  for other values of parameters a bounce is not possible.

The SVW HL cosmology is a bit more complicated as there are  additional terms  in the analogs of Friedmann equations. There exist bouncing solutions for some values of parameters of the theory, however a range of parameters that  lead  only to singular solutions is  wider than in the detailed balance scenario. One very interesting special case includes two  centres, with a saddle between them (corresponding to three finite critical points). If one takes into account  dynamical change of state parameter,  which is much more realistic scenario, it might  be possible to go from one oscillating   solution to another bouncing solution. The problem is that the existence of such solutions depends on the values of coupling constants $\sigma_3$ and $\sigma_4$ and their physical interpretation still remains an open question.

Moreover, in both these formulations, bouncing non-singular solutions exist only in case of a non-flat universe $K\neq0$. Otherwise the bouncing solutions become singular.

\section{Conclusions}

The obtained cosmological results presented here are promising and suggest there is a possibility to replace the initial cosmological singularity of GR by finite bouncing solutions. However, one must also consider that there are many  problems and contradicting statements in the different formulations and extensions of  HL-type theories.  Aside from that aspect there are  also observational bounds on the existence   of the Ho\v{r}ava-Lifshitz gravity  and  the values of its constants and parameters.

At present, HL-type theories, including the original one  and its extensions, are not  yet ruled out  by observational data. However, there now are tight bounds on some parameters of the theory \cite{EmirGumrukcuoglu:2017cfa} from the  binary neutron star merger GW170817 \cite{LIGOScientific:2017zic}. Therefore, it is possible that   further observational  data  might  either rule out some specific scenarios or the whole model. It is also possible that  some agreement with observations  could provide a better justification for additional  theoretical research as it is still  hoped  that HL gravity could offer a promising cosmological scenario without initial singularity and solve some shortcomings of classical GR, like non-renormalisability and thus problems with quantisation. 

There are several observational bounds on different regions of the Ho\v{r}ava-Lifshitz framework, {\it e.g.} using data from binary pulsars~\cite{Yagi:2013ava,Yagi:2013qpa}, using general cosmological
data~\cite{Dutta:2009jn}  and also bounds in the context of dark energy~\cite{Park:2009zr}.  In the context of dark matter and dark energy there are also  bounds on generally Lorentz violation~\cite{Audren:2014hza, Audren:2013dwa}. There is also quite recent  research  performed  in the effective field theory formalism \cite{Frusciante:2015maa} of the extension HL gravity~\cite{Blas:2009qj}. However, this analysis is reduced to a flat background spacetime, which  limits the overall number of parameters.  

In our  papers \cite{Nilsson:2018knn,Nilsson:2023} we have placed  new bounds on parameters of Ho\v{r}ava-Lifshitz cosmology, in its projectable version with and without imposing detailed balance condition. We found  very interesting results on spatial curvature. Namely the original HL model is well fitted with a positive  non-zero spatial curvature with accuracy  to more than  $3\sigma$, whereas  when we relaxed the detailed balance condition we obtained again  positive non-zero spatial curvature at 1$\sigma$ accuracy. As this analysis also included BAO's, therefore  there is needed further investigation of the curvature  parameter which could possibly  finally exclude some of the HL models.  Anyway those results seem to be fascinating in view of future observation and also somehow demonstrate why an analysis limited to zero spatial curvature is somehow limited. Still non-singular bouncing solutions in HL universe appear only for non-zero spatial curvature, so these two topics are related.

 We have to take into account that most obtained bounds on  the parameters of the HL cosmology are similar to those  in  $\Lambda$CDM model, except the non-zero curvature parameter.
 Of course,  the $\Lambda$CDM model  has still fewer parameters  and from this point of view should be preferred; it also   fits the data well. However, one has to bear in mind also the theoretical aspects of Ho\v{r}ava gravity which make it  a good candidate  for an ultra violet complete theory of gravity. There also   several implication like the possible resolution of the initial cosmological singularity,  so there are still many reasons to keep investigating this model and its extensions. 
\bibliography{HLBounce1.bib}

\end{document}